\documentclass[11pt]{article}

\begin{document}

\title{FLRW UNIVERSES FROM ``WAVE-LIKE" COSMOLOGIES IN $5D$}
\author{J. Ponce de Leon\thanks{E-mail: jponce@upracd.upr.clu.edu}\\ Laboratory of Theoretical Physics, Department of Physics\\ 
University of Puerto Rico, P.O. Box 23343, San Juan, \\ PR 00931, USA} 
\date{March 2003}

\maketitle
\begin{abstract}

We consider the evolution  of a $4D$-universe embedded in a five-dimensional (bulk) world with a large extra dimension and a cosmological constant. The cosmology in $5D$  possesses ``wave-like" character in the sense 
that the metric coefficients in the bulk are functions of the extra coordinate and time in a way similar to a pulse or traveling wave propagating  along the fifth dimension. This assumption is motivated by some recent work presenting the big-bang as a higher dimensional shock wave. We show that this assumption, together with an equation of state for the effective matter quantities in $4D$, allows Einstein's equations to be fully integrated. We then recover the familiar FLRW universes, on the four-dimensional hypersurfaces orthogonal to the extra dimension. Regarding the extra dimension we find that it is {\em growing} in size if the universe is speeding up its expansion. We also get an estimate for the relative change of the extra dimension over time. This  estimate could have important observational implications, notably   for the time variation of rest mass,  electric charge and the gravitational ``constant". Our results extend previous ones in the literature.    
 \end{abstract}



\newpage

\section{Introduction}

The possibility that our universe is embedded in a higher dimensional space has generated a great deal of active interest. In Brane-World  and Space-Time-Matter (STM) theories the usual constraint on Kaluza-Klein models, namely the cylinder condition, is relaxed so the  extra dimensions are not restricted to be  ``small".  Although these theories have different physical motivations for the introduction of a large extra dimension, they share the same working scenario, lead to   the same dynamics in $4D$,  and face the same challenges \cite{equ. STM-Brane}. Among them, to predict observationally testable effects of the extra dimension.

The nontrivial dependence of the spacetime metric on the extra coordinate, allowed in both brane-world and STM, endows solutions of the five-dimensional equations with new intriguing properties. The first important question concerning solutions in $5D$ is to check whether they give back standard four-dimensional results. Next, comes the study of the new physics and predictions coming from the extra non-compact dimension. 

A number of solutions are already known. However, of particular interest are those obtained on simple assumptions that allow a complete and systematic integration of the five-dimensional Einstein's equations. As an illustration, we mention  the following three classes of cosmological solutions. 

(i) Solutions where the metric coefficients admit separation of variables \cite{JPdeL 1}. These have been applied to the discussion of a wide variety of cosmological problems that range from singularities to geodesic motion \cite{Wesson book}-\cite{Seahra 2}. 

 (ii) Solutions where $g_{44}$ is static \cite{Binetruy}. These have been discussed in the context of brane models where our universe is a domain wall in a five-dimensional anti-de Sitter spacetime
 \cite{RS2}-\cite{Dadhich}, and in STM for vanishing cosmological constant \cite{LiuWesson}. 

(iii) Solutions where  $g_{00}$ is independent of the extra coordinate. These have been used to study the possibility of variable physical ``constants" in the context of brane-world models \cite{NewVarG} and, a subclass of them, in the analysis of singularities in STM \cite{Fukui Seahra and Wesson}.

However, there is  another simple class of models that allow straightforward integration of the field equations. We refer here to the case where the metric functions have a simple functional dependence of time and the extra coordinate similar to  that in traveling waves or pulses propagating in the fifth dimension.  These ``wave-like" cosmological models were originally investigated by Liu and Wesson \cite{LiuWesson2} and more recently by Wesson, Liu and Seahra \cite{LiuWesson3} in the context of STM. They considered the special class of solutions resulting from the following two conditions. 
\begin{enumerate}
\item The effective matter quantities satisfy the isothermal equation of state $p = \gamma \rho$.
\item The time metric coefficient $g_{00}$ and the expansion factor of the space-sections are related by a power law. 
\end{enumerate} 
Under these conditions, they showed that for both the matter-dominated $(\gamma = 0)$ and radiation-dominated  $(\gamma = 1/3)$ eras of the universe, the $4D$ spacetime defined by hypersurfaces of the $5D$ metric are just the same as those of the standard Friedmann-Robertson-Walker models of general relativity. 

This paper is motivated by the works of Liu and Wesson \cite{LiuWesson2} and Wesson, Liu and Seahra \cite{LiuWesson3}. We believe that wave-like cosmologies represent an elegant class of solutions. But, unfortunately, they have not been investigated enough in the literature. Our aim in this paper is to remedy in part this situation.

Firstly, we show that wave-like cosmologies can be fully integrated in $5D$ without introducing additional assumptions, except for an equation of state. In particular the second condition mentioned above is not needed.  Besides, for future applications in brane theory,  we include  a non-vanishing five-dimensional  cosmological constant in the discussion. 

Secondly, we show that on the four-dimensional hypersurfaces orthogonal to the extra dimension, we recover the familiar FLRW universes, for an arbitrary value of $\gamma$ (not only for $\gamma = 0$ or $\gamma = 1/3$), regardless of the specific choice of parameters in the solution in $5D$.

We also obtain some ``new" physics regarding the extra dimension. Namely, we find that the ``size" of the extra dimension (the metric coefficient of the extra dimension) is related to the expansion factor of the space-sections through the Hubble parameter. An interesting prediction of the model is that, although the extra dimension is small today, it is {\em growing} in size if the universe is speeding up its expansion. The opposite also holds, the size of the extra dimension  is decreasing if the universe is speeding down its expansion. Another significant feature is  that  the relative ``speed" of  change of the extra dimension is determined by the Hubble and deceleration parameters.  This last feature has important observational implications for theories in more than four dimensions that predict the time-variation of some quantities usually considered as constants, among them the gravitational ``constant" $G$ as well as the rest mass and charge of particles.

The paper is organized as follows. In Section 2 we present the ``cosmological" equations in $5D$. In Section 3 we obtain the equation that provides the cosmological evolution of  the wave-like model. In Section 4 we introduce the effective matter. This is a little more involved here than in STM, due to the non-vanishing energy-momentum tensor in $5D$. In Sections 5 and 6 we discuss the solution in the $5D$-bulk and the $4D$-spacetime, respectively. Finally, in Section 7 we present a summary and conclusions. 

\section{Field equations}

In cosmological applications the metric is commonly taken  in the form

\begin{equation}
\label{cosmological metric}
d{\cal{S}}^2 = n^2(t,y)dt^2 - a^2(t,y)\left[\frac{dr^2}{(1 - kr^2)} + r^2(d\theta^2 + \sin^2\theta d\phi^2)\right] + \epsilon \Phi^2(t, y)dy^2,
\end{equation}
where $k = 0, +1, -1$ and $t, r, \theta$ and $\phi$ are the usual coordinates for a spacetime with spherically symmetric spatial sections. We adopt signature $(+ - - - )$ for spacetime and the factor $\epsilon $ can be $- 1$ or $+ 1$ depending on whether the extra dimension is spacelike or timelike, respectively.

The corresponding field equations in $5D$ are 
\begin{equation}
\label{equations in 5D}
G_{AB} = k_{(5)}^2 {^{(5)}T}_{AB},
\end{equation}
where $k_{(5)}^2$ is a constant introduced for dimensional considerations, ${^{(5)}T}_{AB}$ is the energy-momentum tensor in $5D$ and  the non-vanishing components of the Einstein tensor $G_{AB}$ are

\begin{equation}
\label{G 00}
G_{0}^{0} = \frac{3}{n^2}\left(\frac{{\dot{a}}^2}{a^2} + \frac{\dot{a}\dot{\Phi}}{a \Phi}\right) + \frac{ 3 \epsilon}{\Phi^2}\left(\frac{a''}{a} + \frac{{a'}^2}{a^2} - \frac{a' \Phi'}{a \Phi}\right) + \frac{3 k}{a^2},
\end{equation}

\begin{eqnarray}
\label{G 11}
G^{1}_{1} = G^{2}_{2} = G^{3}_{3} &=& \frac{1}{n^2}\left[\frac{\ddot{\Phi}}{\Phi} + \frac{2\ddot{a}}{a} + \frac{\dot{\Phi}}{\Phi}\left(\frac{2 \dot{a}}{a} - \frac{\dot{n}}{n}\right) + \frac{\dot{a}}{a}\left(\frac{\dot{a}}{a} - \frac{2 \dot{n}}{n}\right)\right] + \nonumber \\
& & \frac{\epsilon}{\Phi^2}\left[\frac{2 a''}{a} + \frac{n''}{n} + \frac{a'}{a}\left(\frac{a'}{a} + \frac{2 n'}{n}\right) - \frac{\Phi'}{\Phi}\left(\frac{2a'}{a} + \frac{n'}{n}\right)\right] + \frac{k}{a^2},
\end{eqnarray}
\begin{equation}
\label{G zero four}
G^{0}_{4} = \frac{3}{n^2}\left(\frac{{\dot{a}}'}{a} - \frac{\dot{a} n'}{a n} - \frac{a' \dot{\Phi}}{a \Phi}\right),
\end{equation}
and 
\begin{equation}
\label{G 44}
G_{4}^{4} = \frac{3}{n^2}\left(\frac{\ddot{a}}{a} + \frac{{\dot{a}}^2}{a^2} - \frac{\dot{a}\dot{n}}{a n}\right) + \frac{ 3 \epsilon}{\Phi^2}\left(\frac{{a'}^2}{a^2} + \frac{a' n'}{a n}\right) + \frac{3 k}{a^2}. 
\end{equation}
Here a dot and a prime denote partial derivatives with respect to $t$ and $y$, respectively. 

Introducing the function \cite{Binetruy}
\begin{equation}
\label{F}
F(t,y) = k a^2 + \frac{(\dot{a}a)^2}{n^2} + \epsilon \frac{(a' a)^2}{\Phi^2},
\end{equation}
which is a first integral of the field equations, we get  
\begin{equation}
\label{F prime}
F' = \frac{2a' a^3}{3}k_{(5)}^2 {^{(5)}T}^{0}_{0},
\end{equation}
and
\begin{equation}
\label{F dot}
\dot{F} = \frac{2\dot{a} a^3}{3}k_{(5)}^2 {^{(5)}T}^{4}_{4}.
\end{equation}

Bearing in mind a net cosmological constant in $4D$, in what follows we will assume that the five-dimensional energy-momentum tensor has the form
\begin{equation}
\label{AdS}
{^{(5)}T}_{AB} =  \Lambda_{(5)}g_{AB}, 
\end{equation}
where $\Lambda_{(5)}$ is the cosmological constant in the bulk. We will see that, in this case the effective energy-momentum tensor in $4D$ is conserved. We notice that $\Lambda_{(5)}$ can be (i) positive as in the usual de Sitter $( dS_{5} )$ solution, (ii) negative as in the brane-world scenarios where our spacetime is identified with a singular hypersurface (or $3$-brane) embedded in an $AdS_{5}$ bulk, or (iii) zero as in STM where the matter in $4D$ is interpreted as an effect of the geometry in $5D$. 

\section{The wave-like model}
Now, following Liu and Wesson we assume that the metric coefficients in (\ref{cosmological metric}) are ``wave-like" functions of the argument $(t - \lambda y)$:
\begin{equation}
\label{wave-like}
n = n(t - \lambda y), \;\;\;a = a(t - \lambda y), \;\;\; \Phi = \Phi(t - \lambda y),
\end{equation}
where $\lambda$ is a constant. It is clear that these functions have the same values for coordinates $t$ and $y$ that satisfy the relation $t - \lambda y = const$. Therefore we can say that the values of the metric functions are propagated\footnote{The wave-like model requires $\lambda \neq 0$, otherwise the $5D$ metric is independent of $y$ like in the classical Kaluza-Klein theory with cylindricity.} along the extra dimension $y$ with  ``speed" $\lambda^{- 1} = dy/dt$.

Now, from $G^{0}_{4} = 0$ we get
\begin{equation}
\label{a dot in terms of n and Phi}
\dot{a} = \alpha n \Phi,
\end{equation}
where $\alpha$ is a constant of integration. Substituting this into (\ref{F}) we obtain
\begin{equation}
\label{definition of f}
F = a^2\left(k + \alpha^2 \Phi^2 + \epsilon \lambda^2 \alpha^2 n^2\right) \equiv a^2 f^2.
\end{equation}
The auxiliary function $f$ satisfies the equation 
\begin{equation}
a f \frac{df}{da} + f^2 = \frac{a^2}{3}k_{(5)}^2 \Lambda_{(5)},
\end{equation}
which follows from (\ref{F dot}), (\ref{AdS}) and (\ref{definition of f}). Integrating we get 
\begin{equation}
f^2 = \frac{\beta \alpha^2}{a^2} + \frac{1}{6}a^2 k_{(5)}^2 \Lambda_{(5)},
\end{equation}
where $\beta$ is a constant of integration. Consequently, 
\begin{equation}
\label{Phi}
\Phi^2 = - \epsilon \lambda^2 n^2 - \frac{k}{\alpha^2} + \frac{\beta}{a^2} + \frac{a^2}{6 \alpha^2} k_{(5)}^2 \Lambda_{(5)},
\end{equation}
and  from (\ref{a dot in terms of n and Phi})
\begin{equation}
\label{relation between a and n}
\left(\frac{\dot{a}}{n}\right)^2 + k = - \epsilon \lambda^2 \alpha^2 n^2 + \frac{\beta \alpha^2}{a^2} + \frac{a^2}{6}k_{(5)}^2 \Lambda_{(5)}.
\end{equation}
After some manipulations one can verify that the remaining field equation $G_{1}^{1} = G_{2}^{2} = G_{3}^{3} = k_{(5)}^2 \Lambda_{(5)}$ is identically satisfied. 

Thus, the complete specification of the solution requires the consideration of some physics, or a simplifying mathematical assumption, to determine $ \dot{a}$ (or $n$). Then, from (\ref{relation between a and n}) we find $n$ (or $a$). Finally, the function $\Phi$ is given by  (\ref{Phi}). 

It is not difficult to show that the solution, thus specified, depends only on  {\em two} parameters; namely $\beta$ and $\lambda$. Indeed, let us consider the scale transformation  $y = \xi \bar{y}$, where $\xi$ is some constant. It leads to $\lambda = \bar{\lambda}/\xi$ and $\Phi = \bar{\Phi}/\xi$. If we choose   $\xi = \alpha$, then we get rid of $\alpha$  in the set of equations (\ref{a dot in terms of n and Phi}), (\ref{Phi}) and (\ref{relation between a and n}), provided we retune the constant $\beta$ as  $\beta \alpha^2$ ($\bar{\beta} = \beta \alpha^2$).   Consequently, one can always set $\alpha = 1$, without loss of generality. 

In what follows, however we will keep  $\alpha$ arbitrary and $\lambda \neq 0$. The case $\lambda = 0$, for which the metric is independent of $y$,  is examined in Section 7. The physical interpretation of $\beta$ in terms of the so-called black or Weyl radiation is discussed in Section 5.

\section{Effective matter}

The simplest way of completing the above system of equations is making some assumption on the metric functions for example $n = 1$, or  some more general relation $n = n(a)$. The same can also be done for $\Phi$. 

However, we consider here a more physical approach. Namely, we are concerned about the ``effective" matter induced in $4D$. We show that if we assume an equation of state, for the effective quantities, then the problem becomes fully determined, i.e., we do not have to assume anything else. 

The interpretation of the five-dimensional field equations, in terms of four-dimensional quantities, is provided by the fact that, for an {\em arbitrary} five-dimensional metric,  the $15$ equations (\ref{equations in 5D}) in $5D$ can be split up into three parts. Namely, a set of $10$ equations which looks effectively as the field equations in $4D$ with  an effective  energy-momentum tensor. A set of $4$ equations which resembles Maxwell equations with sources, and an equation for the scalar field $\Phi$ \cite{EMT}.

In absence of off-diagonal terms $(g_{4\mu} = 0)$ the dimensional reduction of the five-dimensional equations  is particularly simple. The usual assumption  is that our spacetime is orthogonal to the extra dimension. Thus we introduce the normal unit ($n_{A}n^{A} = \epsilon$) vector, orthogonal to hypersurfaces $y = constant$,  
\begin{equation}
n^A = \frac{\delta^{A}_{4}}{\Phi}\;  , \;\;\;\;\;  n_{A}= (0, 0, 0, 0, \epsilon \Phi).
\end{equation}
Consequently,  the physical metric coincides with the spacetime part of (\ref{cosmological metric}). Following the usual procedure, we define the energy-momentum tensor in four-dimensions (on the hypersurfaces $y = const$) through the effective Einstein field equations in $4D$, namely 
\begin{eqnarray}
\label{4D Einstein with T and K}
{^{(4)}G}_{\alpha\beta} &=& \frac{2}{3}k_{(5)}^2\left[^{(5)}T_{\alpha\beta} + (^{(5)}T^{4}_{4} - \frac{1}{4}{^{(5)}T})g_{\alpha\beta}\right] -\nonumber \\
& &\epsilon\left(K_{\alpha\lambda}K^{\lambda}_{\beta} - K_{\lambda}^{\lambda}K_{\alpha\beta}\right) + \frac{\epsilon}{2} g_{\alpha\beta}\left(K_{\lambda\rho}K^{\lambda\rho} - (K^{\lambda}_{\lambda})^2 \right) - \epsilon E_{\alpha\beta}, 
\end{eqnarray}
where $K_{\mu\nu}$ is the extrinsic curvature 
\begin{equation}
\label{extrinsic curvature}
K_{\alpha\beta} = \frac{1}{2}{\cal{L}}_{n}g_{\alpha\beta} = \frac{1}{2\Phi}\frac{\partial{g_{\alpha\beta}}}{\partial y},\;\;\; K_{A4} = 0,
\end{equation}
and $E_{\mu\nu}$ is the projection of the bulk Weyl tensor ${^{(5)}C}_{ABCD}$ orthogonal to ${\hat{n}}^A$, i.e., ``parallel" to spacetime, viz.,
\begin{eqnarray}
\label{Weyl Tensor}
E_{\alpha\beta} &=& {^{(5)}C}_{\alpha A \beta B}n^An^B\nonumber \\
&=& - \frac{1}{\Phi}\frac{\partial K_{\alpha\beta}}{\partial y} + K_{\alpha\rho}K^{\rho}_{\beta} - \epsilon \frac{\Phi_{\alpha;\beta}}{\Phi} - \epsilon \frac{k^{2}_{(5)}}{3}\left[{^{(5)}T}_{\alpha\beta} + ({^{(5)}T}^{4}_{4} - \frac{1}{2}{^{(5)}T})g_{\alpha\beta}\right].
\end{eqnarray}
The first term on the r.h.s. of (\ref{4D Einstein with T and K}) yields
\begin{equation}
T_{\mu\nu}^{(\Lambda)} = \frac{1}{2}k_{(5)}^2\Lambda_{(5)}g_{\mu\nu} \equiv \Lambda g_{\mu\nu},
\end{equation}
where we have used (\ref{AdS}) and $\Lambda = k_{(5)}^2\Lambda_{(5)}/2$ is the effective cosmological constant in $4D$. The rest of the  terms on the r.h.s. of (\ref{4D Einstein with T and K}) represent the effective  mater, viz.,  
\begin{equation}
8 \pi GT_{\mu\nu}^{(eff)} \equiv  - \epsilon\left(K_{\mu\lambda}K^{\lambda}_{\nu} - K_{\lambda}^{\lambda}K_{\mu\nu}\right) + \frac{\epsilon}{2} g_{\mu\nu}\left(K_{\lambda\rho}K^{\lambda\rho} - (K^{\lambda}_{\lambda})^2 \right) - \epsilon E_{\mu\nu}. 
\end{equation}
In cosmological problems, the effective matter is commonly assumed to be  a perfect fluid
\begin{equation}
T_{\mu\nu}^{(eff)} = (\rho_{eff} + p_{eff}) u_{\mu}u_{\nu} - p_{eff} g_{\mu\nu}.
\end{equation}
It is clear that this interpretation is not unique. For example, we could  assume that the effective matter is the superposition of several fluids with distinct equations of state. What we are doing here, since we want to recover known results, is a ``modest" extension of four-dimensional general relativity. Besides, only the effective (or total) quantities have observational consequences.

As a consequence of the contracted Bianchi identities in $4D$, ${^{(4)}G}^{\mu}_{\nu; \mu} = 0$,  and (\ref{AdS}), the effective energy-momentum tensor satisfies the standard general relativity conservation equations, viz., 
\begin{equation}
\nabla^{\mu}T_{\mu\nu}^{(eff)} = 0.
\end{equation}
If in the bulk there were scalar and/or other fields, then this would be no longer true, in general.

Since $G_{01} = 0$, it follows that $u_{1} = 0$. Hence the effective perfect fluid is ``at rest" in the frame given by (\ref{cosmological metric}). Thus, for the case under consideration, the  appropriate equations are, 
\begin{eqnarray}
\label{effective quantities}
8 \pi G \rho_{eff} + \Lambda  &=& \frac{3}{n^2}\left(\frac{\dot{a}}{a}\right)^2 + \frac{3 k}{a^2},\nonumber \\
8 \pi G p_{eff} - \Lambda &=& - \frac{1}{n^2}\left[\frac{2 \ddot{a}}{a} + \frac{\dot{a}}{a}\left(\frac{\dot{a}}{a} - \frac{2\dot{n}}{n}\right)\right] - \frac{k}{a^2},
\end{eqnarray}
It is important to emphasize here that the effective matter content of the spacetime will be the same whether we interpret it as induced matter, as in STM, or as the ``total" matter in a ${\bf Z}_2$ symmetric brane universe. This is a consequence of the identification of the tensor $P_{\mu\nu}$ of STM with the energy-momentum tensor on the brane in brane theory \cite{equ. STM-Brane}, \cite{EMT}.

It is useful to introduce  the ``proper" time $\tau$, as
\begin{equation}
\label{proper time}
d\tau = n dt.
\end{equation}
In terms of which the expressions for matter density and pressure become simpler. Namely,
\begin{eqnarray}
\label{density and pressure in terms of proper time}
8 \pi G \rho_{eff} + \Lambda  &=& \frac{3}{a^2}\left(\frac{d a}{d \tau}\right)^2 + \frac{3 k}{a^2},\nonumber \\
8 \pi G p_{eff} - \Lambda &=& - \frac{2}{a}\left(\frac{d^2 a}{d\tau^2}\right) - \frac{1}{a^2}\left(\frac{da}{d\tau}\right)^2 - \frac{k}{a^2}.
\end{eqnarray}
We now assume that the effective matter quantities satisfy the isothermal equation of state\footnote{Negative values of $\gamma$ are also allowed in the study of inflationary models of our universe, which require violation of the ``strong" energy condition, viz.,  $(\rho_{eff} + p_{eff}) < 0 $.}, viz.,
\begin{equation}
\label{equation of state}
p_{eff} = \gamma \rho_{eff}, \;\;\;\  0 \leq \gamma \leq 1.
\end{equation}
This provides a differential  equation for $a$, viz.,
\begin{equation}
\frac{2}{a}\left(\frac{d^2 a}{d \tau^2}\right) + \frac{(3 \gamma + 1)}{a^2}\left(\frac{d a}{d \tau}\right)^2 + (3 \gamma + 1)\frac{k}{a^2} = (\gamma + 1)\Lambda,
\end{equation}
whose first integral is 

\begin{equation}
\label{first integral}
\left(\frac{da}{d\tau}\right)^2 = \frac{C}{a^{3\gamma + 1}} - k + \frac{a^2}{3}\Lambda,
\end{equation}
where $C$ is a constant of integration related to the effective matter in $4D$. Thus, solving this equation and using (\ref{proper time}) we obtain the solution as a function of the argument $(t - \lambda y)$.

\section{The solution in the bulk}

Thus, the problem of finding the wave-like cosmological metrics in $5D$ becomes totally specified by the equation of state (\ref{equation of state}). Collecting results, the  metric functions in the five-dimensional metric (\ref{cosmological metric}) are given by
\begin{eqnarray}
\label{n and Phi}
n^2 &=& \frac{\epsilon}{\lambda^2 \alpha^2}\left(\frac{\beta \alpha^2}{a^2} - \frac{C}{a^{(3\gamma + 1)}}  - \frac{a^2}{6} \Lambda\right),\nonumber \\
\Phi^2 &=& \frac{1}{\alpha^2}\left(\frac{C}{ a^{(3\gamma + 1)}} - {k} + \frac{a^2}{3} \Lambda\right),
\end{eqnarray}
where $a$ is the solution of the equation
\begin{equation}
\label{a dot}
{\dot{a}}^2 = \frac{\epsilon}{\lambda^2 \alpha^2}\left(\frac{C}{a^{3\gamma + 1}} - k + \frac{a^2}{3}\Lambda\right)\left(\frac{\beta \alpha^2}{a^2} - \frac{C}{a^{(3\gamma + 1)}}  - \frac{a^2}{6} \Lambda\right).
\end{equation} 
We note that the parameter $\beta$ is related to the so-called Weyl or black radiation. Indeed, substituting (\ref{a dot in terms of n and Phi}) and (\ref{n and Phi}) into (\ref{Weyl Tensor}) we obtain
\begin{equation}
\label{black radiation}
8 \pi G \rho_{Weyl} = - \epsilon E_{0}^{0} = \frac{3 \beta \alpha^2}{a^4}, \;\;\; p_{Weyl} = \frac{1}{3}\rho_{Weyl}, 
\end{equation}
where $8 \pi G p_{Weyl} = \epsilon E^{1}_{1} = \epsilon E^{2}_{2} = \epsilon E^{3}_{3}$. 
Thus setting $\beta = 0$ is equivalent to eliminating the contribution coming from the free gravitational field. 

\medskip

Some particular solutions to (\ref{a dot}), namely those with $\beta = 0$, $k = 0$ and $\Lambda = 0$ where discussed by Liu and Wesson.  Although they make the additional assumption $n = a^{-(3 \gamma + 1)/2}$, which as we see here is unnecessary. 

\medskip 

We note that  the asymptotic behavior of the solution for $a \rightarrow 0$ is independent of $k$ and $\Lambda$. In this limit, the positiveness of $\Phi^2$ requires $C > 0$. The particular case $\beta = 0$ requires  the extra dimension to  be spacelike $(\epsilon = -1)$. For $\beta \neq 0$, it can be either spacelike or timelike. For $\gamma = 1/3$, which corresponds to a radiation-dominated era, $n^2 > 0$ demands $\epsilon(\beta \alpha^2 - C) > 0.$ Besides, from (\ref{a dot}), with $(\gamma = 1/3)$, it follows that both the matter and the  Weyl radiation (i.e., $C$ and $\beta$)   contribute to  the dynamics, viz.,

\begin{equation}
\label{approx sol for radiation dom era}
a^2 \approx \left(\frac{9 \epsilon C(\beta \alpha^2 - C)}{\lambda^2 \alpha^2}\right)^{1/3}(t - \lambda y)^{2/3}.
\end{equation}
For $\gamma < 1/3$, which includes dust $(\gamma = 0)$, the Weyl contribution dominates over the matter and ${\dot{a}}^2 \sim  (\epsilon \beta C/\lambda^2)a^{- 3(\gamma +1)}$ with $\epsilon \beta > 0$, as $a \rightarrow 0$. 
For $\gamma > 1/3$, which includes the stiff equation of state $p_{eff} = \rho_{eff}$, it is the opposite. Namely,  in this limit the effective matter dominates over the Weyl term and ${\dot{a}}^2 \sim  - \epsilon C^2/a^{(6 \gamma + 2)}$. This becomes possible only if the extra dimension is spacelike\footnote{If the extra dimension is timelike, then $\beta$ must be positive and $a$ is bounded from bellow. For example, for $\gamma = 1$, $a^2 \geq a^2_{min} \approx C/\beta \alpha^2$.} $(\epsilon = -1).$

At late times, the specific details of the solution in $5D$  for $a >> 1$ do depend on $k$, $\beta$, $\Lambda$ and the signature of the extra dimension.

\section{The solution in $4D$: FLRW universe}

Since (\ref{a dot}) depends on a number of parameters, it is clear that the general solution in $5D$ is very rich and complex. What is amazing here is that the physical scenario  in $4D$ is always the same regardless of the specific choice of the $5D$ parameters $\beta$, $\lambda$ and signature of the extra dimension. Namely,  

\begin{equation}
\label{FLRW metric}
d{\cal{S}}^2 = d\tau_{0}^2 - a_{0}^{2}(\tau)\left[\frac{dr^2}{(1 - kr^2)} + r^2(d\theta^2 + \sin^2\theta d\phi^2)\right],
\end{equation} 
where the evolution of the scale factor is given by the usual equation in FLRW universes 

\begin{equation}
\label{FLRW1}
\left(\frac{da_{0}}{d\tau_{0}}\right)^2 = \frac{C}{a_{0}^{3\gamma + 1}} - k + \frac{a_{0}^{2}}{3}\Lambda.
\end{equation}
Here the subscript $0$ for $\tau$ and $a$ means that these functions are evaluated at some $y = const$.
The corresponding energy density is given by  
\begin{equation}
\label{FLRW2}
8 \pi G \rho_{eff} = \frac{3C}{a_{0}^{3(\gamma + 1)}}.
\end{equation}
Since the solutions to these equations can be found in many textbooks, we are not going to discuss them here. 

In addition to the FLRW universes, we obtain some specific expressions for the extra dimension. Indeed, let us  notice that\footnote{In what follows we omit the subscript $0$.} 
\begin{equation}
\Phi^2 = \frac{1}{\alpha^2} \left(\frac{da}{d\tau}\right)^2.
\end{equation}
Consequently the ``size" of the extra dimension is related to  the matter in $4D$ as 

\begin{equation}
\label{effects of matter of Phi}
\alpha \frac{d\Phi}{d\tau} = \left[\Lambda - 4 \pi G (3\gamma + 1)\rho_{eff}\right]\frac{a}{3} = \frac{d^2a}{d\tau^2}.
\end{equation}
Thus for the rate of change of $\Phi$ we get

\begin{equation}
\label{rate of change of Phi}
\frac{1}{\Phi}\frac{d \Phi}{d\tau} = -  q H,
\end{equation}
where $H= (a_{\tau}/a)$ and $q = - (a a_{\tau \tau}/a_{\tau}^2)$ are the Hubble and ``deceleration" $4D$ parameters, respectively. According to modern observations, the universe is expanding with an acceleration, so that the parameter $q$ is, roughly, $- 0.5 \pm 0.2$. Assuming  $H = h \times 10^{-10} yr^{-1}$ we come to the estimate

\begin{equation}
\label{estimate}
\frac{1}{\Phi}\frac{d\Phi}{d\tau} \approx (3.5 \pm 1.4)\times 10^{-10} yr^{-1},
\end{equation}
where we have taken $h = 0.7$ \cite{Melnikov2}. Also for the dimensionless ratio $\Phi/a$ we have

\begin{equation}
\frac{\Phi}{a}  = \alpha^{-1} \times h\times 10^{- 10}.
\end{equation}
 We also note that the above relations hold for the three cases, $k = - 1, 0, + 1 $ and arbitrary cosmological constant.

\section{Summary and conclusions}

In this work we have discussed cosmologies in five-dimensional Kaluza-Klein theory where the metric coefficients have a wave-like behavior, in the sense that they depend on the single variable $(t - \lambda y)$. The non-vanishing  parameter  $\lambda $ is the one that permits the universe to have diverse equations of state during its evolution. Namely, in the limit $\lambda \rightarrow 0$ from (\ref{relation between a and n})  we get

\begin{equation}
\left(\frac{da}{d\tau}\right)^2  + k = \frac{\beta \alpha^2}{a^2} + \frac{a^2}{3}\Lambda.
\end{equation}
Then, from (\ref{density and pressure in terms of proper time}) it follows that 

\begin{equation}
8 \pi G \rho_{eff} = 8 \pi G \rho_{Weyl} = \frac{3 \beta \alpha^2}{a^4}.
\end{equation}
Thus for $\lambda = 0$ there is only (Weyl) radiation, while for $\lambda \neq 0$ other equations of state are possible as we have seen in (\ref{FLRW1}) and (\ref{FLRW2}). 

We have shown that an equation of state, for the effective matter quantities,  completely specifies  wavelike cosmologies in $5D$.  We note that the introduction of induced matter is not reduced to a trivial isolation of  the ``extra" $5D$-terms in the r.h.s. of the $5D$-equations. If we had proceed in this way, we would have obtained $k_{(4)}^2\Lambda_{(5)}$ (or $2 \Lambda$) in (\ref{effective quantities}) instead of the correct term $\Lambda = k_{(4)}^2\Lambda_{(5)}/2 $.

The specific form of the solution in $5D$ depends on the choice of the parameters $(\beta, C, k, \Lambda_{(5)})$ and the signature of the extra dimension. In four dimensions (on the hypersurfaces $y = const$) these cosmologies are just the same as those of the standard FLRW universes with $k = 0, + 1, -1$. 

Our model also predicts the development  of the extra dimension. Equations (\ref{FLRW2}) and (\ref{effects of matter of Phi})  show how  the dynamics of $\Phi$ is influenced by the effective matter.   On the basis of our model we can reach some conclusions.

(i) Although $\Phi$ is small today, it is {\em growing} in size if the universe is speeding up its expansion. The opposite also holds, the size of $\Phi$ is decreasing is the universe is speeding down its expansion. 

(ii) The relative change of $\Phi$ is determined by the Hubble and deceleration parameters as shown in (\ref{rate of change of Phi}). 

(iii) At any time during  the evolution\footnote{The parameter $\alpha$ can be taken as $\alpha = 1$, without loss of generality.} $(\alpha \Phi) = H a$. 

These conclusions are general in the sense that they are the same regardless of the details of the model in $4D$, ie., the value of $\gamma$, $k$ and $\Lambda$.

 The estimate given by (\ref{estimate}), almost certainly cannot  be independently tested. However, it  could have important observational implications. This is because  the  ratio $(\dot{\Phi}/\Phi)$ appears in different contexts, notably  in expressions concerning the variation of rest mass \cite{defofmass},  electric charge \cite{massandcharge} and variation of the gravitational ``constant" $G$ \cite{Melnikov3},\cite{Melnikov1}.  

Indeed, in the Randall-Sundrum brane-world scenario and other non-compact Kaluza-Klein theories,  the motion of test particles is higher-dimensional in nature. In other words, all test particles travel on five-dimensional geodesics but observers, who are bounded to spacetime, have access only to the $4D$ part of the trajectory. In general, the effective rest mass measured in $4D$ changes as the test particle  travels on $5D$ geodesics\footnote{The general, invariant equations for the change of mass are given in \cite{defofmass}.}. The total change consists of two parts, one of them is induced by the non-trivial dependence of the metric on the extra coordinate $(\partial g_{\mu\nu}/\partial y \neq 0)$ and the other part is due to $\dot{\Phi}/\Phi$. Even in the simplest situation, where the metric does not depend on the extra coordinate, but only on time, $m_{0}$ the effective rest mass  in $4D$ of a  massless particle in $5D$ would change as

\begin{equation}
\frac{1}{m_{0}}\frac{dm_{0}}{d\tau} = - \frac{1}{\Phi}\frac{d\Phi}{d\tau},
\end{equation}
where we have used equation (25) in Ref. \cite{defofmass}. Similarly, the variation of $\Phi$ induces a  change in the  electric charge, and consequently in the fine structure constant \cite{massandcharge}.  

Regarding the time-variation of $G$, it is remarkable that in different  models with extra dimensions the ratio $(\dot{G}/G)$  is found  to be proportional to $(\dot{\Phi}/\Phi)$ \cite{NewVarG},\cite{Melnikov3}. At this point we have to mention that the specific value of $(\dot{\Phi}/\Phi)$ depends on the cosmological model. For example, for the cosmologies with separable metric coefficients mentioned in the Introduction $(\dot{\Phi}/\Phi) = (1 + q)H$, instead of (\ref{rate of change of Phi}).  Consequently, the measurement of quantities like $(\dot{m_{0}}/m_{0})$ and $(\dot{G}/G)$ will give the opportunity to test different models for compatibility with observational data.

 For completeness we mention another important aspect of these cosmologies pointed out by Liu and Wesson for the case with $\Lambda = 0$.  It concerns the nature of the big-bang which occurs at $a = 0$. In this limit the scale factor  is given by (\ref{approx sol for radiation dom era}), $a \sim (t - \lambda y)^{1/3}$. If observers have different values of $y$, they will experience the big-bang as having occurred at different times $t$.  Thus they will measure different values for the age of the universe. 

\paragraph{Acknowledgments}I would like to thank V.N. Melnikov, J.F. Nieves and Paul S. Wesson for comments on the ratio $\dot{\Phi}/\Phi$.

\end{document}